\pdfoutput=1

\documentclass[11pt]{article}

\usepackage[final]{acl}

\usepackage{times}
\usepackage{latexsym}

\usepackage[T1]{fontenc}

\usepackage[utf8]{inputenc}

\usepackage{microtype}

\usepackage{inconsolata}

\usepackage{graphicx}

\usepackage{kotex}
\usepackage{multirow}
\usepackage{graphicx}
\usepackage{balance}
\usepackage{soul}
\usepackage{bbm}
\usepackage{adjustbox}
\usepackage{tabularx}
\usepackage{booktabs}
\usepackage{arydshln}
\usepackage{amsmath}

\usepackage{caption}
\title{RRADistill: Distilling LLMs' Passage Ranking Ability \\for Long-Tail Queries Document Re-Ranking on a Search Engine}

\author{
 \textbf{Nayoung Choi\textsuperscript{2$\dagger$*}},
 \textbf{Youngjune Lee\textsuperscript{1$\dagger$}},
 \textbf{Gyu-Hwung Cho\textsuperscript{1}},
 \textbf{Haeyu Jeong\textsuperscript{1}},
\\
 \textbf{Jungmin Kong\textsuperscript{1}},
 \textbf{Saehun Kim\textsuperscript{1}},
 \textbf{Keunchan Park\textsuperscript{1}},
 \textbf{Sarah Cho\textsuperscript{1}},
 \textbf{Inchang Jeong\textsuperscript{1}},
\\
 \textbf{Gyohee Nam\textsuperscript{1}},
 \textbf{Sunghoon Han\textsuperscript{1}},
 \textbf{Wonil Yang\textsuperscript{1}}
 \textbf{and Jaeho Choi\textsuperscript{1}}
\\
 \textsuperscript{1}Naver Corporation,
 \textsuperscript{2}Emory University
\\
 \texttt{
   nayoung.choi@emory.edu, youngjune.lee93@navercorp.com
 }
}

\begin{document}
\maketitle
\def\thefootnote{$\dagger$}\footnotetext{Both authors contributed equally to this research.}
\def\thefootnote{*}\footnotetext{Work done while at Naver.}
\begin{abstract}
Large Language Models (LLMs) excel at understanding the semantic relationships between queries and documents, even with lengthy and complex long-tail queries. These queries are challenging for feedback-based rankings due to sparse user engagement and limited feedback, making LLMs' relevance ranking ability highly valuable. However, the large size and slow inference of LLMs necessitate the development of smaller, more efficient Small Language Models (SLMs). Recently, integrating ranking label generation into distillation techniques has become crucial, but existing methods underutilize LLMs' capabilities and are cumbersome. Our research, \textbf{RRADistill} (\textbf{R}e-\textbf{R}anking \textbf{A}bility \textbf{Distill}ation), propose an efficient label generation pipeline and novel SLM training methods for both encoder and decoder models. We introduce an encoder-based method using a Term Control Layer to capture term matching signals and a decoder-based model with a ranking layer for enhanced understanding. Experimental results including A/B testing on NAVER, South Korea's leading search platform, demonstrate effectiveness of our approach in re-ranking for long-tail queries.
\end{abstract}

\section{Introduction}
\begin{figure}[t!]
  \centering
  \includegraphics[width=\linewidth]{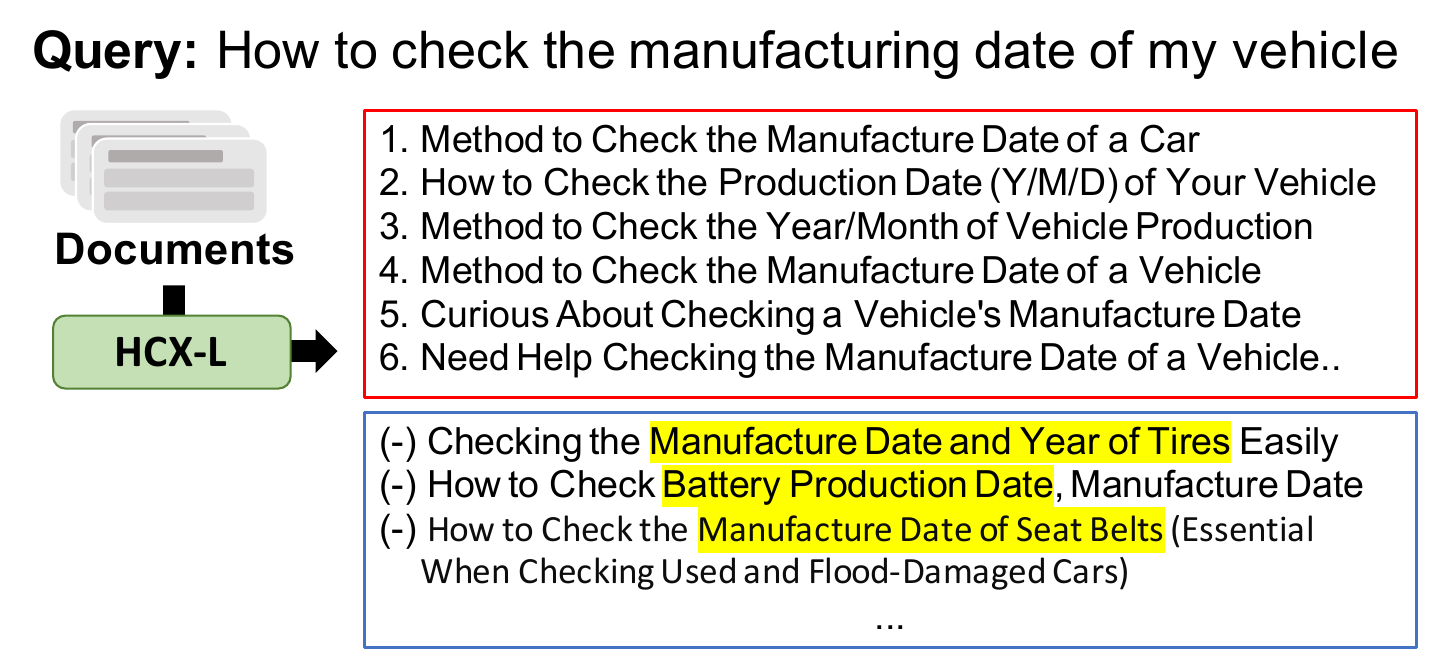}
  \caption{An example of zero-shot inference by HyperCLOVA X (HCX) on retrieved documents. \textcolor{red}{The red box} shows HCX's ranked output; \textcolor{blue}{the blue box} shows excluded documents. Irrelevant parts of excluded documents are \hl{highlighted in yellow}. The original was in Korean, but translated to English.}
  \vspace{-10pt}
  \label{fig:hcx_zeroshot}
\end{figure}

Providing relevant search results to enhance user satisfaction is crucial in search engines. To achieve this, various research efforts have been conducted on different search tasks, such as retrieval and ranking \cite{splade, colbert, nogueira2019passage, toter, taxoindex}. Large Language Models (LLMs), such as ChatGPT \cite{chatgpt} and GPT-4 \cite{openai2024gpt4}, have shown remarkable potential across diverse search tasks, including query rewriting \cite{mao-etal-2023-large, dhole2024genqrensemble}, query and document expansions \cite{wang2023query2doc, mackie2023grf, ma2023pretraining}. As LLMs advance in complex tasks, they also show potential for passage ranking, which involves understanding relationships between queries and multiple documents. Recent studies \cite{qin2023large, ma2023zeroshot, zhuang2024yes, pradeep2023rankvicuna} show that instruction-tuned LLMs like GPT-4, Flan-T5 \cite{JMLR:v25:23-0870} and Vicuna \cite{vicuna2023} effectively handle passage ranking in zero-shot settings. Motivated by these studies, we also conducted zero-shot re-ranking with our in-house LLMs, HyperCLOVA X (HCX) \cite{yoo2024hyperclova} which comprises Large (HCX-L) and Small (HCX-S) variants. We put a query and a list of document snippet texts to HCX, using a Korean translation of list-wise prompt from RankGPT \cite{sun-etal-2023-chatgpt}. We found that HCX-L effectively returns document identifiers in the desired order, excluding low-relevance ones, as depicted in Figure \ref{fig:hcx_zeroshot}. Unlike short-head queries, which are popular and short like keywords, long-tail queries are complex and involve longer, specific phrases (\ref{appendix:longtail_query}). These queries benefit more from relevance than feedback, and semantic than syntactic matching due to their rich semantic content but lack of user feedback. Thus, LLMs' ability to rank complex long-tail queries by relevance is highly valuable.

However, the slow inference speed challenges the direct use of LLMs in search engine. To address this, we trained a much smaller Language Model (SLM) to retain HCX-L's ranking ability. This area of research is known as knowledge distillation \cite{dcd,derrd, kang2021topology,ccd,hetcomp}, where the knowledge of large models is transferred to smaller models to improve efficiency in information retrieval. Our approach builds on similar LLM distillation methods, such as those used in RankGPT and TWOLAR \cite{TWOLAR}. This involves two stages: 1) Generating ranking label using LLMs, and 2) Training the SLM ranker. When generating ranking label with LLMs, previous studies utilize the list-wise permutation generation method, which inputs a query and a set of documents to LLM and receives an ordered list of document identifiers. However, these approach require sliding windows, which infer multiple times on partial lists due to the prompt length constraints of LLMs, causing a burden. Moreover, previous studies viewed the \textit{missing} phenomenon as a problem, where LLMs fail to include all input documents in the output. However, we observed that in most cases, excluded documents due to \textit{missing} are significantly irrelevant to the query, making it a valuable signal. Consequently, we reframed the \textit{missing} and highlighted its impact. We developed our own label generation pipeline to address these issues, including two key techniques: 1) Pre-rank to filter documents, retaining only those effective to train SLM rankers and bypass the sliding window, 2) Consider \textit{missing} as useful signals, and utilize excluded documents as hard negatives, to train SLM rankers. Our pipeline speeds up labeling and provides compact yet effective training data.

In training SLM rankers, we explored both BERT \cite{devlin-etal-2019-bert} and GPT \cite{gpt2} styles, incorporating our training techniques. For BERT ranker, we integrate a term control layer into the training process to utilize specific term matching signals. For GPT ranker, we developed techniques to effectively utilize classification (whether \textit{relevant} or \textit{irrelevant}) and reasoning (rationale for \textit{relevant} or \textit{irrelevant}) during training, with a light-weight ranking layer. Both rankers incorporate additional training layers, but only specific parts of the model architecture (Encoder plus a classification head for BERT, Decoder plus a dense layer for GPT) are utilized during inference, reducing the burden for service applications. 

In this paper, we provide various experiments on our BERT ranker (RRA-BERT) and GPT ranker (RRA-GPT), trained to mimic LLMs' relevance ranking, aiming to improve long-tail search quality. We tested the effectiveness of our methodology through rigorous online and offline evaluations. 


\section{Methodology}
\subsection{Label Generation with LLMs}
\label{sec:label_generation}

\begin{figure*}[t!]
\centering
  \includegraphics[width=1.0\linewidth]{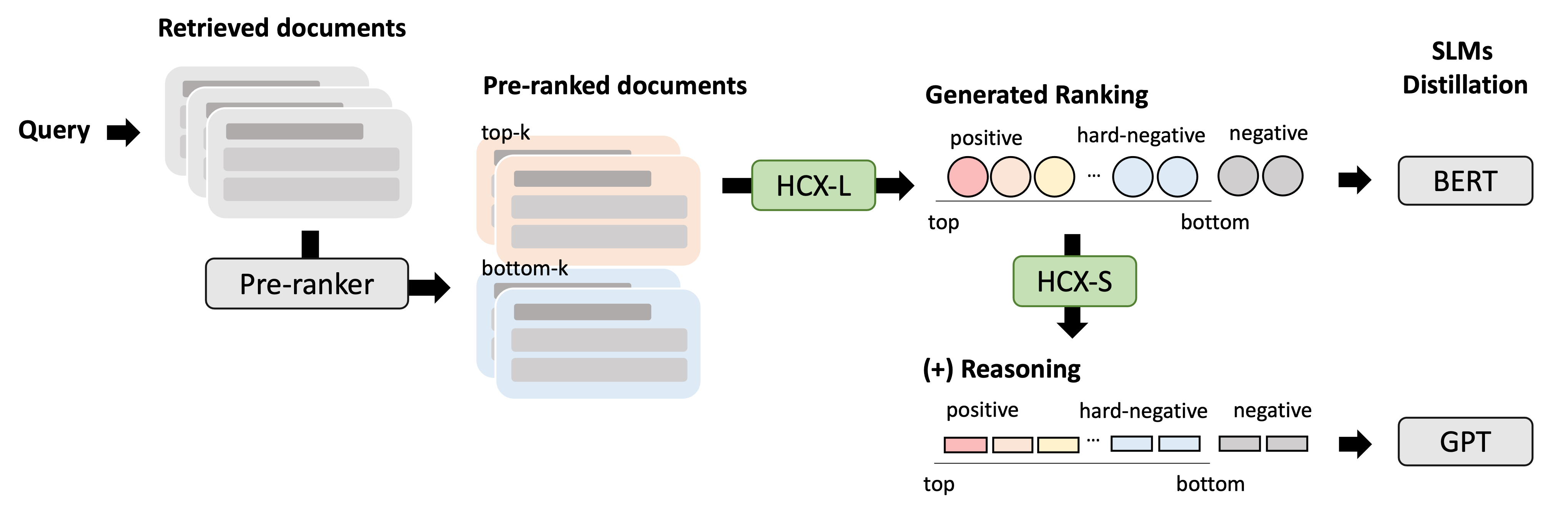}
  \caption{The overview of our label generation pipeline. Negatives are randomly selected from documents totally unrelated to the query.}
  \label{fig:overview}
\end{figure*}

First, we sampled 7,000 long-tail queries from NAVER search logs based on length, complex phrasing and frequency criteria, as detailed in \ref{appendix:longtail_query}. Then, we retrieved 50 documents per query with multiple retrievers of NAVER search engine. Given a query $q$ and retrieved documents $D = [d_1, d_2, ..., d_n]$, we ranked $D$ using our pre-ranker. From pre-ranker ($Ranker_{\text{pre}}$), we obtained the top 10 ($D_{\text{top10}}$) and bottom 10 ($D_{\text{bottom10}}$) documents, and labeled them with HCX-L \cite{yoo2024hyperclova} in a list-wise manner. All document inputs are snippet texts. Figure \ref{fig:overview} depicts the overview of our label generation pipeline with LLM. Further details, including the pre-ranker (\ref{appendix:preranking}) are described in \ref{appendix:label_generation_detail}.
\begin{equation}
\begin{aligned}
    D' = Ranker_{\text{pre}}(D) \\
    D_{\text{top10}} = D'[:10]; \; D_{\text{bottom10}} = D'[-10:] \\
    D_{\text{pre}} = D_{\text{top10}} \cup D_{\text{bottom10}} \\
    D_{\text{ranked}} = HCX_{\text{L}}(Prompt(q, D_{\text{pre}})) \\
    D_{\text{excluded}} = D_{\text{pre}} \setminus D_{\text{ranked}}
\end{aligned}
\end{equation}
\noindent Previous study \cite{sun-etal-2023-chatgpt} has highlighted the \textit{missing} phenomenon, where LLMs rank only part of the input list, suggesting its frequency varies depending on LLMs. HCX-L also exhibited frequent missing occurrences. However, as shown in Figure \ref{fig:hcx_zeroshot}, we observed that in most cases only relevant documents were included in the output ($D_{\text{ranked}}$), excluding documents with significantly low relevance to the query ($D_{\text{excluded}}$). Hence, we reframed the \textit{missing} as a valuable signal, leveraging excluded documents as hard negatives. Through comparison experiments in Section \ref{sec:bert_ablation}, training with and without excluded documents, we demonstrated the usefulness of the missing signal. For GPT ranker training, we generated reasoning, which is the rationale for why $q$ and $d$ is \textit{relevant} or \textit{irrelevant}, as described in \ref{appendix:reasoning}.

\subsection{BERT-sytle Distillation: RRA-BERT}
\label{sec:bert}
\begin{figure}[ht!]
\centering
  \includegraphics[width=\linewidth]{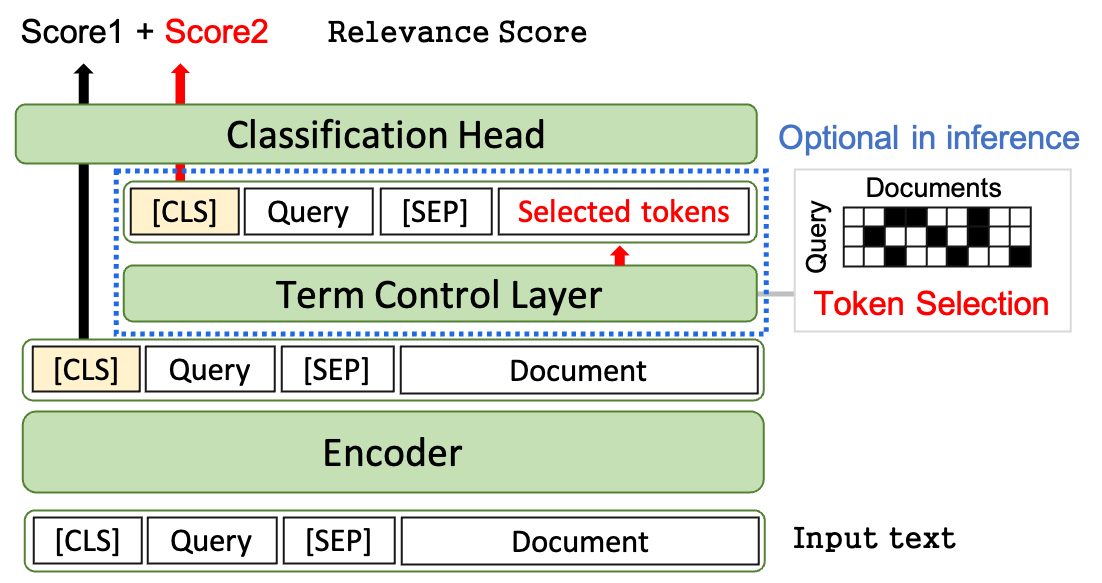}
  \caption{RRA-BERT: The [SEP] token distinguishes the query and the document. The Term Control Layer can be omitted during inference.}
  \label{fig:bert}
\end{figure}

Lengthy and complex queries require attention not only to the overall semantic information but also to 
specific terms that are particularly noteworthy within the query. To address this problem, we propose a novel training approach designed to inject term matching signals between queries and documents as hints into dense representations to effectively enhance performance. BERT-style distillation consists of three components: (1) \textbf{Token Selection (TS)} method to select tokens from the document matched to the query. (2) \textbf{Term Control Layer (TCL)} that utilizes information of selected tokens as hints for training. (3) \textbf{Optimization} that effectively combines semantic and term information in the training process. The overall structure of BERT-style distillation is in Figure \ref{fig:bert}.

\subsubsection{\textbf{Token Selection}}
\label{sec:token_selection}
We propose a Token Selection (TS) method that captures terms matching signals between the query and the documents as hints, allowing the model to consider both the overall and specific semantics. We select top-$k$ document tokens that match each query token. The process involves identifying and selecting matched tokens $T_{q, d}$ within document $d$ for a given query $q$ using word embeddings ($Embedding_{\text{word}}$), as follows.
\begin{equation}
\begin{aligned}
E_{q} = Embedding_{\text{word}}(Tokenizer(q)) \\\
E_{d} = Embedding_{\text{word}}(Tokenizer(d)) \\
Sim_{q,d} = E_{q}E^T_{d} \\
T_{q, d} = \{\text{Top}_k(Sim_{q,d}[i, :]) | i = 1, \ldots, n_q\}
\end{aligned}
\end{equation}
where $n_q$ is the number of tokens in $q$. In this process, we select $k$ $\times$ $n_q$ tokens from the document, while excluding duplicate tokens.

\subsubsection{\textbf{Term Control Layer}}
\label{sec:tcl}
We propose Term Control Layer (TCL) that effectively integrates the term matching signals into the overall training process. Unlike the overall semantic score (Score1 in Figure \ref{fig:bert}), TCL utilizes only selected document tokens $T_{q,d}$ to focus on specific tokens. We designed TCL with a multi-head self-attention \cite{10.5555/3295222.3295349} mechanism using the last hidden states of encoder as the input, enabling the aggregation of information from each token. The corresponding formula is as follows.
\begin{equation}
\label{eq:TCL}
\begin{aligned}
H_{\text{T}} = Concat(h_{\text{[CLS]}}, h_{q}, h_{\text{[SEP]}}, h_{T_{q,d}})\\
TCL(H_{\text{T}})= Attention_{\text{multi-head}}(H_{\text{T}})
\end{aligned}
\end{equation}
where $h_{\text{[CLS]}}$, $h_{\text{q}}$, $h_{\text{[SEP]}}$ and $h_{T_{q,d}}$ represent the last hidden states of $[CLS]$, the query, $[SEP]$ and $T_{q,d}$, respectively. The number of attention heads is 8.

\subsubsection{\textbf{Optimization}}
\label{sec:optimization}
To compute the basic ranking score $s_{base}$ (\texttt{Score1} in Figure \ref{fig:bert}), we input the representation $h_{\text{[CLS]}}$ to the classification head ($CLF
_{\text{head}}$). Then, we calculate the score $s_{TCL}$  (\textcolor{red}{\texttt{Score2}} in Figure \ref{fig:bert}) in the same manner, using the TCL-derived $h_{\text{[CLS]}}$ which is TCL$(H_{\text{T}})_{\text{[CLS]}}$ in equation \ref{eq:TCL}. Both calculations share the same $CLF_{\text{head}}$. The final relevance score $s$ is obtained as follows. 
\begin{equation}
\begin{aligned}
\label{eq:bert_optimization}
 s_{\text{base}} = CLF_{\text{head}}(h_{\text{[CLS]}}) \\
 s_{\text{TCL}} = CLF_{\text{head}}(TCL(H_{\text{T}})_{\text{[CLS]}}) \\
 s = s_{\text{base}} + \alpha * s_{\text{TCL}}
 \end{aligned}
 \end{equation}
where $\alpha$ controls the effects of TCL. Finally, given $S$, a list of outputs $s$ for $n$ documents, we compute the training loss using RankNet \cite{ranknetloss}, a pairwise loss function.
\begin{equation}
\begin{aligned}
 L = L_{RankNet}(S); \; \text{where } S = [s_1, s_2, \ldots, s_n]
 \end{aligned}
 \label{eq:ranknet_loss}
 \end{equation}
Our relevance score computation is designed to effectively capture overall semantics with $s_{\text{base}}$, while focusing on specific matching terms with TCL, represented by $s_{\text{TCL}}$. Also, this design improves $s_{\text{base}}$ by naturally integrating the term signal into its computation. Therefore, we can remove the TCL during inference to reduce the deployment burden without degrading performance. A detailed analysis is provided in Section \ref{sec:bert_ablation}.

\subsection{GPT-sytle Distillation: RRA-GPT}
\label{sec:gpt}
Existing GPT and T5 \cite{DBLP:journals/corr/abs-1910-10683} rankers primarily use decoder output as a relevance score, calculating from the decoder logits of either the first generated token or the generated target tokens (e.g., \textit{Yes} or \textit{No}). To optimize relevance scores, training tasks typically fall into two types: classification, as implemented in MonoT5 \cite{nogueira2020document}, and ranking, exemplified by RankT5 \cite{zhuang2022rankt5}, TWOLAR, and RankGPT. Furthermore, ExaRanker \cite{10.1145/3539618.3592067}, a T5-based ranker, enhances ranking performance by training to generate explanations for \textit{relevant} or \textit{irrelevant}. In this paper, we explored which task combinations help GPT ranker training and whether reasoning enhances the ranking performance. 

\begin{figure}[h!]
\centering
\includegraphics[width=\linewidth]{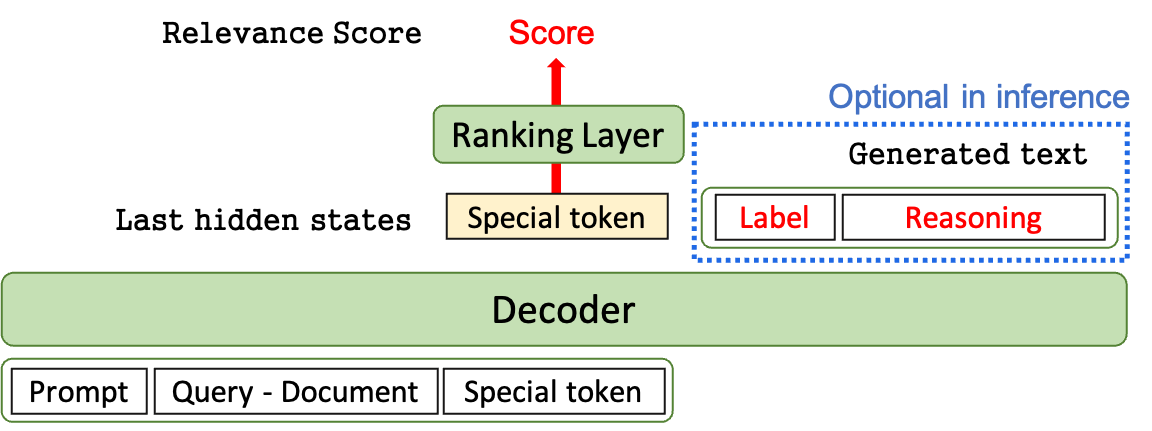}
  \caption{RRA-GPT: The special token here is \texttt{<|Response|>}. The label and reasoning generation can be omitted during inference.}
  \label{fig:gpt}
\end{figure}

In previous studies \cite{sun-etal-2023-chatgpt, zhang2023rankwithoutgpt, pradeep2023rankzephyr}, it was surprising to see that despite GPT's much larger parameter size, its performance often matched or fell behind that of BERT and T5 rankers, which suggests GPT lacks a dedicated input encoding(=understanding) module. To address this, we enhanced GPT ranker with a dense layer, which we call a ranking layer. Selecting the appropriate input for this layer is critical, akin to [CLS] token in BERT, to effectively represent the $(q,d)$ relationship. We tested token embeddings from both the input text and the generated texts, \texttt{<|Response|>} and \texttt{<|Reason|>} respectively, as input to the ranking layer. Our final GPT ranker is depicted in Figure \ref{fig:gpt}. 

\begin{figure}[h!]
\centering  \includegraphics[width=\linewidth]{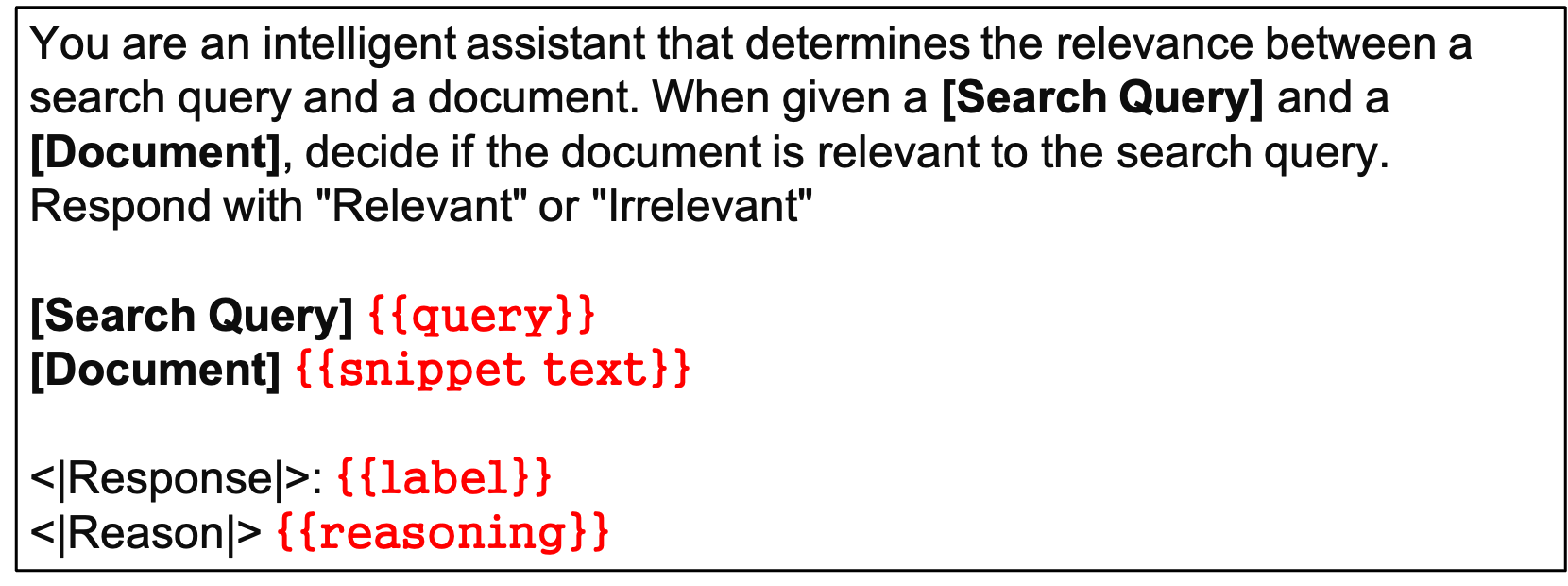}
  \caption{Training data format: Replace \textcolor{red}{the red sections} \small{$\{\{query\}\}, \{\{snippet\text{ }text\}\}, \{\{label\}\}, \{\{reasoning\}\}$} as needed.}
  \label{fig:prompt}
\end{figure}

During training, we jointly train relevance ranking, label generation (\texttt{<|Relevant|>} or \texttt{<|Irrelevant|>}), and reasoning. In inference, only the ranking layer is used, without any label generation. The format of our training prompt is shown in Figure \ref{fig:prompt}, but during inference, the prompt includes up to the "\texttt{<|Response|>}:" part only.

We used significantly small GPT as a backbone for ranker. The backbone has been instruction-tuned, including a task that generates label for a $(q,d)$ pair as either \textit{relevant} or \textit{irrelevant}. We added a total of four special tokens: \texttt{<|Relevant|>}, \texttt{<|Irrelevant|>}, \texttt{<|Response|>} and \texttt{<|Reason|>}, to distinguish them from tokens in $q$ and $d$ during model training. Each special token is initialized with corresponding token embedding; e.g., \texttt{<|Relevant|>} is initialized with "Relevant".

\subsubsection{\textbf{Ranking layer}}
\label{sec:ranking_layer}
The relevance score $s$ for a ${(q,d)}$ pair is obtained through a dense layer as follows.
\begin{equation}
\resizebox{\linewidth}{!}{$
\begin{aligned}
X = Tokenizer(Prompt(q, d)) \\
H = Decoder(X) \\
h_{\texttt{<|Resp.|>}} = H[-1, i]; \text{ where } i \text{ is the index of } \texttt{<|Response|>} \text{ in } X \\
s = Dense(h_{\texttt{<|Resp.|>}}); \text{ where } d_{\text{in}}=d_{\text{hidden}}, d_{\text{out}}=1
\end{aligned}
$}
\end{equation}
\noindent where $H$ refers to all layers of hidden states, and $h_{\texttt{<|Resp.|>}}$ is the last hidden state of a special token. The \textit{Dense} layer has an input dimension equal to the model's hidden size and outputs a single relevance score. Given a list of outputs $s$ for $n$ documents, where $S = [s_1, s_2, \ldots, s_n]$, the calculation of the loss $L_{\text{RankNet}}(S')$, where $S' = \text{MinMaxScaling}(S)$, follows the BERT ranker training formula as in Equation \ref{eq:ranknet_loss}.

\subsubsection{\textbf{Ranking with classification \& reasoning}}
We propose leveraging the decoder's generative abilities for ranking layer training. By simultaneously training the decoder to generate labels and reasoning, we enhance the ranking layer. Label generation, akin to classification, is as follows.
\begin{equation}
\resizebox{\linewidth}{!}{$
\begin{aligned}
\label{eq:gpt_rel_score}
z = LM_{\text{head}}(Decoder(X)) \\
z_{\text{rel}} = z[k]; \text{ where } k \text{ is the token id of } \texttt{<|Relevant|>} \\
z_{\text{irrel}} = z[j]; \text{ where } j \text{ is the token id of } \texttt{<|Irrelevant|>} \\
p_{\text{rel}} = \frac{e^{z_{\text{rel}}}}{e^{z_{\text{rel}}} + e^{z_{\text{irrel}}}}, \quad p_{\text{irrel}} = \frac{e^{z_{\text{irrel}}}}{e^{z_{\text{rel}}} + e^{z_{\text{irrel}}}} \\
L_{\text{clf}} = -[y \log(p_{\text{rel}}) + (1-y) \log(p_{\text{irrel}})]
\end{aligned}
$}
\end{equation}
\noindent The $LM_{\text{head}}$ takes input from hidden layers and generates token probabilities over vocabulary auto-regressively. The logit $z$ of the first generated token is always the logit of either \texttt{<|Relevant|>} or \texttt{<|Irrelevant|>}, as training progresses. To infer the class, it is calculated as $1$ if $p_{\text{rel}} > p_{\text{irrel}}$ else $0$ (1=relevant, 0=irrelevant). The final loss including the generation loss is as follows.
\begin{equation}
\resizebox{\linewidth}{!}{$
\label{eq:loss}
\begin{aligned}
X = Tokenizer(Prompt(q, d, label[, reasoning])) \\
L_{\text{gen}} = -\sum_{t=1}^{\text{len}(X)} \log p_{\text{\footnotesize{model}}}(x_t | x_1, x_2, \ldots, x_{t-1}) \\
L = L_{\text{gen}} + L_{\text{RankNet}} + L_{\text{clf}}
\end{aligned}
$}
\end{equation}
\noindent where $p_{\text{\footnotesize{model}}}$ represents the probability that GPT generates the token $x_t$ given the preceding tokens. 

\section{Experiment}
\label{sec:experiment}
We tested the effectiveness of our label generation and training method, which leverages BERT and GPT structures, alongside following baselines, HCX-L \cite{yoo2024hyperclova}, BM25 \cite{robertson2009probabilisticbm25}, MonoBERT \cite{nogueira2019passage}, MonoT5 \cite{nogueira2020document} and RankGPT (\citealt{sun-etal-2023-chatgpt}). MonoBERT and MonoT5 utilize our labeled dataset for training. RankGPT's training labels are generated using sliding windows, without our pre-ranking process. While exact parameter sizes of backbones are undisclosed, they follow the order: T5 (small) $<$ BERT $<<$ GPT $<$ T5 (large), all of which are below 1 billion parameters. 
For evaluation, we used our custom NAVER testset along with Korean-translated public testsets for passage re-ranking: MS MARCO \cite{bajaj2018ms}, MIRACL \cite{10.1162/tacl_a_00595}, DL19 \cite{DBLP:journals/corr/abs-2003-07820}, and DL20 \cite{DBLP:journals/corr/abs-2102-07662}. More detailed settings about dataset, metrics, baselines, and implementation are outlined in \ref{appendix:settings}.

\begin{table*}[t]
\caption{Performance comparison. Best scores per dataset are in \textbf{bold}, with second best scores \underline{underlined}.}
\label{tab:performance_all}
\centering
\begin{adjustbox}{width=\linewidth}

\begin{tabular}{l|rr|rr|rr|rr|rr}
\hline \hline
\multicolumn{1}{r|}{\multirow{2}{*}{}} & \multicolumn{2}{c|}{\textbf{NAVER}}                                                                                                    & \multicolumn{2}{c|}{\textbf{MS MARCO}}                                                                                                 & \multicolumn{2}{c|}{\textbf{MIRACL}}                                                                                                 & \multicolumn{2}{c|}{\textbf{DL19}}                                                                                                     & \multicolumn{2}{c}{\textbf{DL20}}                                                                                                      \\
\multicolumn{1}{r|}{}                  & \multicolumn{1}{c}{nDCG@5}                                    & \multicolumn{1}{c|}{nDCG@10}                                  & \multicolumn{1}{c}{nDCG@5}                                    & \multicolumn{1}{c|}{nDCG@10}                                  & \multicolumn{1}{c}{nDCG@5}                                    & \multicolumn{1}{c|}{nDCG@10}                                & \multicolumn{1}{c}{nDCG@5}                                    & \multicolumn{1}{c|}{nDCG@10}                                  & \multicolumn{1}{c}{nDCG@5}                                    & \multicolumn{1}{c}{nDCG@10}                                   \\ \hline
\textbf{BM25} & 0.427 & 0.520 & 0.418 & 0.524 & 0.473  & 0.568 & 0.350 & 0.396 & 0.277 & 0.284 \\
\textbf{BERT} (naive) & 0.535     & 0.655   & 0.492 &  0.567  &0.671 & 0.740 &  0.584  & 0.601  & 0.388 & 0.419\\
\textbf{GPT} (vanilla) & 0.376 & 0.473 & 0.387 & 0.501 & 0.323 & 0.445 & 0.266 & 0.307 & 0.204 & 0.226 \\
\hline
\textbf{MonoBERT} & 0.639    & 0.757   & \underline{0.533} &  \underline{0.600}  & \textbf{0.696} & \textbf{0.759}& \underline{0.656}  & \textbf{0.662}& \textbf{0.565} &\underline{0.560} \\
\textbf{MonoT5} (large)  & \underline{0.650} & \underline{0.759} & 0.520 & 0.589 & 0.668 &
0.739 & 0.633 & 0.652 & \underline{0.560} & \textbf{0.565} \\ 
\textbf{RankGPT} (bert)  & 0.589 & 0.696 & 0.446 & 0.542 & 0.623 & 0.688 & 0.557 & 0.565 & 0.431 & 0.434 \\ 
\textbf{RankGPT} (gpt)  & 0.432 & 0.535 & 0.363 & 0.487 & 0.284 & 0.415 & 0.295 & 0.327 & 0.180 & 0.201 \\ 
\hline
\textbf{HCX-L} (zero-shot)  & - & - & 0.523 & 0.595 & \underline{0.686} & 0.733 & 0.621 & 0.620 & 0.480 & 0.480 \\ 
\hdashline
\textbf{RRA-BERT} (ours)                            & \multicolumn{1}{r}{\textbf{0.655}} & \multicolumn{1}{r|}{\textbf{0.776}} & \multicolumn{1}{r}{\textbf{0.543}} & \multicolumn{1}{r|}{\textbf{0.607}} & \multicolumn{1}{r}{0.671} & \multicolumn{1}{r|}{\underline{0.743}} & \multicolumn{1}{r}{\textbf{0.667}} & \multicolumn{1}{r|}{\underline{0.658}} & \multicolumn{1}{r}{0.546} & \multicolumn{1}{r}{0.536} \\
\textbf{RRA-GPT} (ours)                             & 0.620 & 0.735 & 0.491 & 0.548 & 0.567 & 0.660 & 0.521 & 0.548 & 0.417 & 0.421 \\ \hline \hline
\end{tabular}

\end{adjustbox}
\end{table*}

\subsection{Results}
Table \ref{tab:performance_all} presents the overall performance. \textbf{RRA-BERT} excels on our long-tail query testset (NAVER) and matches or surpasses other baselines on public testsets, even outperforming the larger MonoT5. Despite its smaller size, \textbf{RRA-GPT} also shows significant improvement, underscoring the effectiveness of our training methods. Moreover, all models trained on our dataset perform comparably to HCX-L, demonstrating the impact of our label generation pipeline. HCX-L relatively underperformed in DL19 and DL20, where many documents necessitated the use of sliding windows (see Table \ref{tab:dataset}). This may be attributed to the finding \cite{zhang2023rankwithoutgpt} that relevant documents are \textit{trapped} in the local block and fail to propagate to the next window. This issue may also clarify why the first-retrieval stage greatly influences LLM re-ranking \cite{sun-etal-2023-chatgpt}, supporting the efficacy of our window-avoiding labeling approach. Our approach has led to a small efficient model that particularly excels at handling complex queries while also performing well with general queries.

\subsubsection{\textbf{RRA-BERT}} 
\label{sec:bert_ablation}
We provide an ablation study and analysis of RRA-BERT addressing three questions: \textbf{(1)} \textbf{The impact of our label generation method}, \textbf{(2)} \textbf{The effectiveness of Token Selection (TS) and Term Control Layer (TCL)}, and \textbf{(3)} \textbf{Inference efficiency}. The experimental results are presented in Table \ref{tab:bert}. \textbf{First}, despite having the same architecture, RankGPT (bert) significantly underperforms MonoBERT and RRA-BERT trained on our dataset, confirming the effectiveness of our label generation pipeline. In addition, an ablation study that excludes LLMs' \textit{missing} documents from the training data (\texttt{w/o missing}) showed a significant drop in performance, demonstrating that the LLMs' \textit{missing} phenomenon is a useful signal. \textbf{Second}, training with TS and TCL (\texttt{w/ TS+TCL}) enhances overall performance while reducing the standard deviation without DL20. TS and TCL contributes to both performance improvement and stable training. \textbf{Third}, training with TS and TCL but removing TCL during inference (\texttt{infer w/o}) did not cause performance degradation. At the same time, removing TCL reduced the inference time by 5.58\%. This indicates that TCL enhances $s_{\text{base}}$ in Equation \ref{eq:bert_optimization} by naturally integrating the term signal into its computation. Therefore, we can remove the TCL during inference without degrading performance. As part of our qualitative evaluation, we provide real-world examples of long-tail query ranking results using RRA-BERT in \ref{appendix:example}. These examples demonstrate that the model effectively handles long-tail query ranking while still benefiting from improved inference efficiency.

\begin{table}[]
\caption{Ablation study on RRA-BERT (nDCG@5 w/ standard deviation). Best scores per dataset are in \textbf{bold}.}
\centering
\begin{adjustbox}{width=\linewidth}
\label{tab:bert}
\begin{tabular}{l|r|r|r|r|r}
\hline \hline
\multicolumn{1}{l|}{}                                                                  & \multicolumn{1}{c|}{\textbf{NAVER}}                           & \multicolumn{1}{c|}{\textbf{\begin{tabular}[c]{@{}c@{}}MS\\ MARCO\end{tabular}}} & \multicolumn{1}{c|}{\textbf{MIRACL}} & \multicolumn{1}{c|}{\textbf{DL19}} & \multicolumn{1}{c}{\textbf{DL20}}  \\
\cline{1-6} 
\texttt{w/o missing}                            & \begin{tabular}[c]{@{}r@{}}0.617\\ \small{($\pm$ 0.006)}\end{tabular} & \begin{tabular}[c]{@{}r@{}}0.532\\ \small{($\pm$ 0.010)}\end{tabular}  & \begin{tabular}[c]{@{}r@{}}0.627\\ \small{($\pm$ 0.041)}\end{tabular}                  &       \begin{tabular}[c]{@{}r@{}}0.616\\ \small{($\pm$ 0.026)}\end{tabular}   &  \begin{tabular}[c]{@{}r@{}}0.537\\ \small{($\pm$ 0.013)}\end{tabular}   \\ \cline{1-6}
\texttt{w/o TS+TCL}                            & \begin{tabular}[c]{@{}r@{}}0.645\\ \small{($\pm$ 0.006)}\end{tabular} & \begin{tabular}[c]{@{}r@{}}0.539\\ \small{($\pm$ 0.006)}\end{tabular}   &   \begin{tabular}[c]{@{}r@{}}0.648\\ \small{($\pm$ 0.045)}\end{tabular}                                     &   \begin{tabular}[c]{@{}r@{}}0.658\\ \small{($\pm$ 0.020)}\end{tabular}     &  \begin{tabular}[c]{@{}r@{}}0.544\\ \small{($\pm$ 0.010)}\end{tabular}  
\\ \cline{1-6}
 \texttt{w/ TS+TCL}                            & \begin{tabular}[c]{@{}r@{}}\textbf{0.655}\\ \small{($\pm$ 0.001)}\end{tabular} & \begin{tabular}[c]{@{}r@{}}\textbf{0.543}\\ \small{($\pm$ 0.004)}\end{tabular}    &  \begin{tabular}[c]{@{}r@{}}0.671\\ \small{($\pm$ 0.019)}\end{tabular}    &  \begin{tabular}[c]{@{}r@{}}\textbf{0.667}\\ \small{($\pm$ 0.010)}\end{tabular}     &    \begin{tabular}[c]{@{}r@{}}0.546\\ \small{($\pm$ 0.024)}\end{tabular}                                \\
\texttt{infer w/o} & \begin{tabular}[c]{@{}r@{}}0.654\\ \small{($\pm$ 0.001)}\end{tabular}       &    \begin{tabular}[c]{@{}r@{}}\textbf{0.543}\\ \small{($\pm$ 0.004)}\end{tabular}       &   \begin{tabular}[c]{@{}r@{}}\textbf{0.674}\\ \small{($\pm$ 0.019)}\end{tabular}   &\begin{tabular}[c]{@{}r@{}}0.664\\ \small{($\pm$ 0.012)}\end{tabular}   &  \begin{tabular}[c]{@{}r@{}}\textbf{0.550}\\ \small{($\pm$ 0.021)}\end{tabular}   \\ 
\hline \hline
\end{tabular}

\end{adjustbox}
\vspace{-5pt}
\end{table}

\begin{figure}[h!]

\centering
\begin{tabular}{cc}
\includegraphics[width=.46\columnwidth]{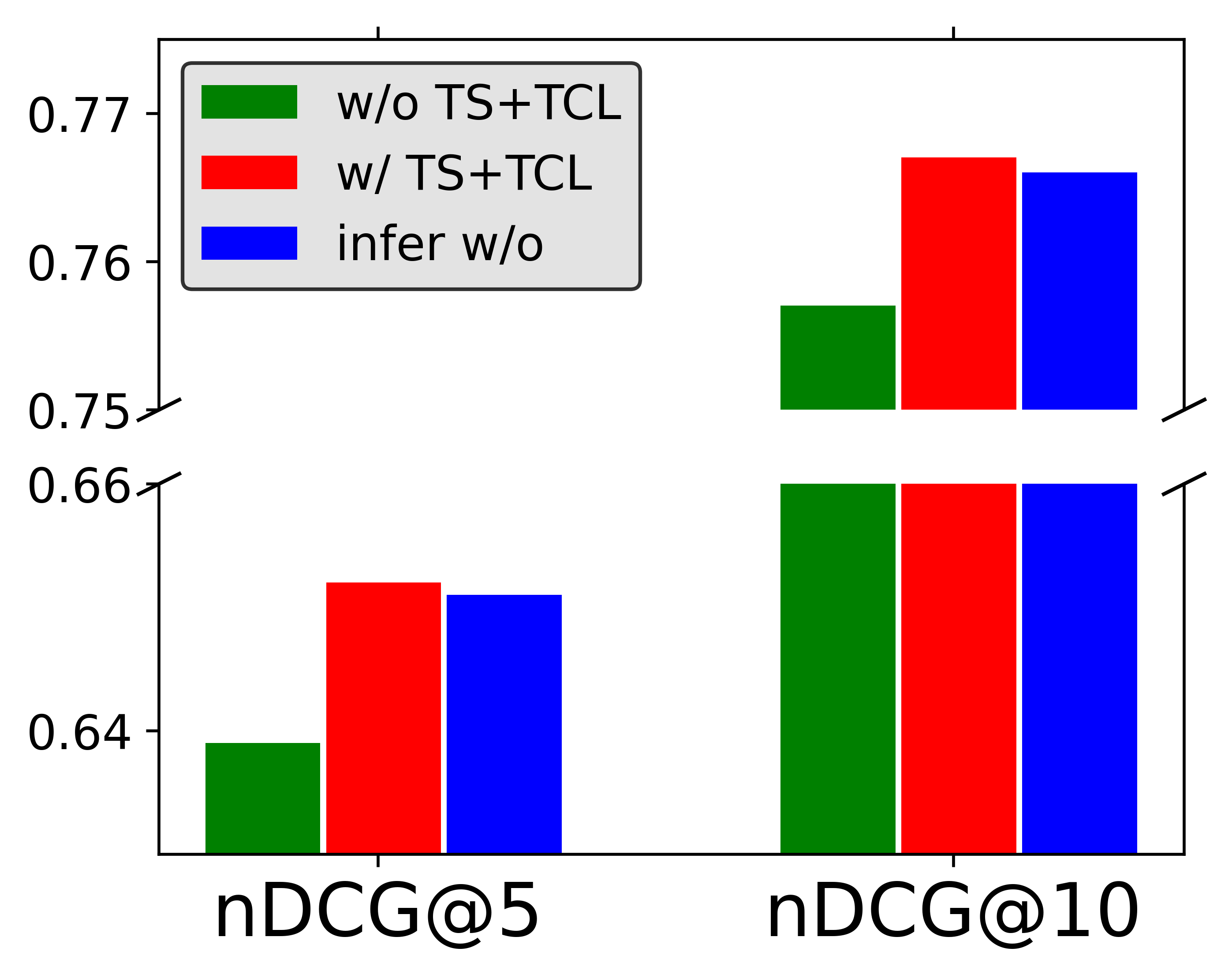}&
\includegraphics[width=.46\columnwidth]{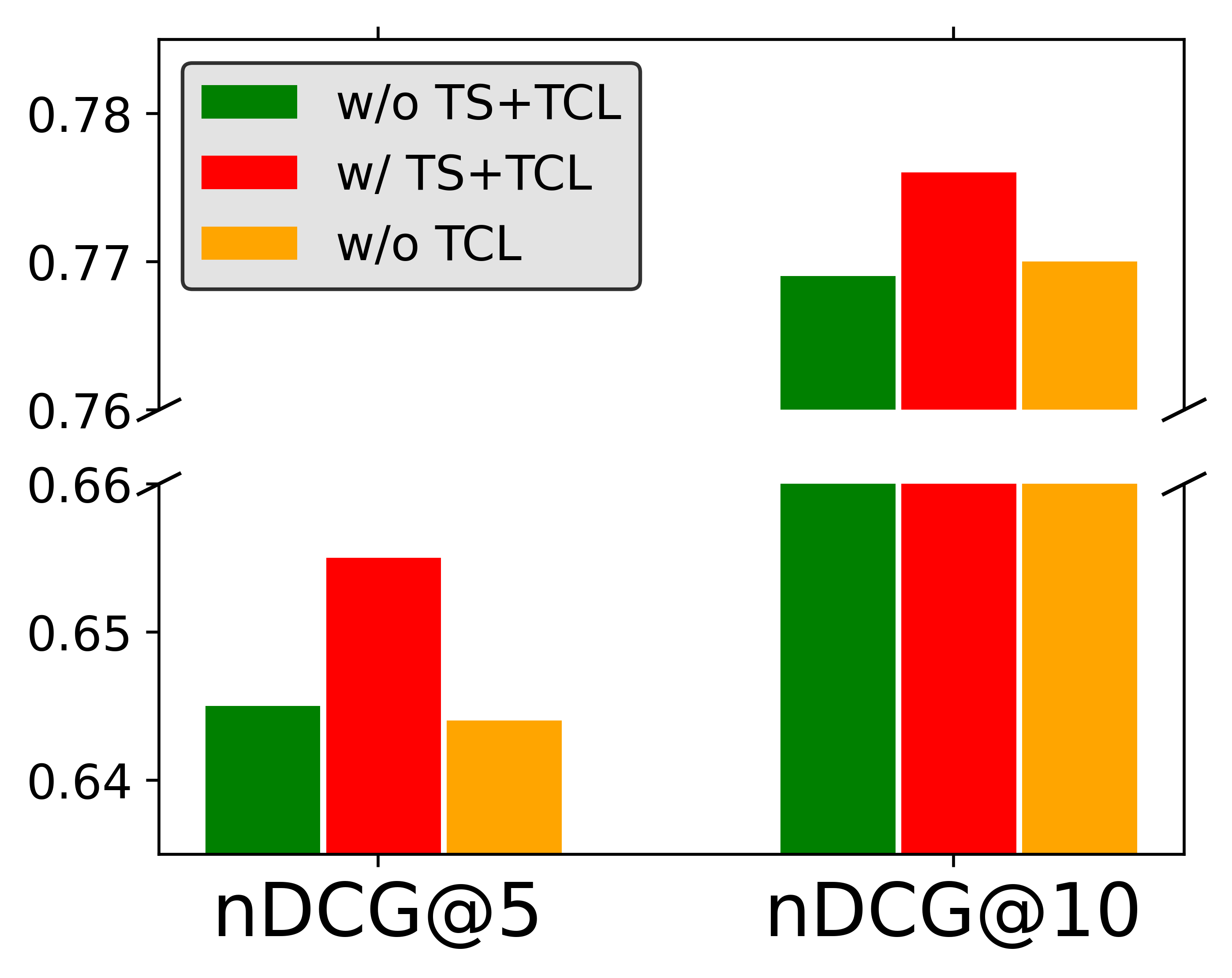}
\end{tabular}
\caption{
Generalizability study of the TS and TCL with MonoBERT (Left) and effectiveness study of TCL using RRA-BERT (Right) on the NAVER testset.
}
\label{fig:TCL}

\end{figure}

We conducted further experiments to validate the generalizability and effectiveness of our approach, as shown in Figure \ref{fig:TCL}. Testing on MonoBERT, we observed performance gains with TS and TCL (\texttt{w/ TS+TCL}) and no decline without TCL during inference (\texttt{infer w/o}), similar to the previous results. This suggests our method reliably enhances performance across ranking models. Additionally, to assess TCL's effectiveness, we input selected document tokens into the encoder, bypassing TCL (\texttt{w/o TCL}), and calculated \textcolor{red}{\texttt{Score2}} in Figure \ref{fig:bert} without TCL. The results showed no improvement over \texttt{w/ TS+TCL}, confirming the effectiveness of TCL in integrating term signals during training.

\subsubsection{\textbf{RRA-GPT}} 
\label{sec:gpt_ablation}
Here we explore three questions: \textbf{(1)} \textbf{The most effective training task combinations}, \textbf{(2)} \textbf{The impact of the ranking layer} and whether input or generated tokens are preferable, and \textbf{(3)} \textbf{The influence of reasoning on ranking training}. The findings are summarized in Table \ref{tab:gpt_performance}. \textbf{First}, models trained with all three tasks: classification (\texttt{clf}), ranking (\texttt{rank}), and generation (\texttt{gen}), showed the best ranking performance. However, adding \texttt{clf} or \texttt{rank} task individually had no effect in our experiments. \textbf{Second}, training a ranking layer was effective, and using the input token embedding \texttt{<|Response|>} (abbr. \texttt{<|Resp.|>}) was better than using the generated one \texttt{<|Reason|>} (abbr. \texttt{<|Rsn.|>}), for the input. \textbf{Third}, adding the ranking layer greatly improves performance with reasoning, while without it, reasoning worsens ranking, a consistent trend across all our experimental units. We found that simply adding reasoning without extra training layer to GPT does not enhance performance. Undoubtedly, even if reasoning does not directly improve ranking performance, it remains valuable for explainable ranking. Given the challenges of interpreting results from neural models, obtaining explanations for the output of the ranker (i.e., why $q$ and $d$ are relevant or not) is crucial.

\begin{table}[h!]
\caption{Ablation study on RRA-GPT (nDCG@10 w/ standard deviation). Best scores per dataset are in \textbf{bold}, with second best scores \underline{underlined}.}
\label{tab:gpt_performance}
\centering
\begin{adjustbox}{width=\linewidth}

\begin{tabular}{lc|r|r|r|r|r}
\hline \hline
\multicolumn{2}{l|}{}                                                                                                                                                                                   & \multicolumn{1}{c|}{\textbf{NAVER}}                                                            & \multicolumn{1}{c|}{\begin{tabular}[c]{@{}c@{}}\textbf{MS} \\ \textbf{MARCO}\end{tabular}}                                                         & \multicolumn{1}{c|}{\textbf{MIRACL}}                                                         & \multicolumn{1}{c|}{\textbf{DL19}}                                                             & \multicolumn{1}{c}{\textbf{DL20}}                                                              \\ \cline{1-7} 
\multicolumn{2}{c|}{\hl{\textbf{w/o ranking layer\protect\footnotemark}}}                                                                                                                                            &                                                                                       &                                                                                       &                                                                                     &                                                                                       &                                                                                       \\
\multicolumn{1}{c}{\textbf{Task}}                                                                               & \textbf{Reasoning}                                                                                      &                                                                                       &                                                                                       &                                                                                     &                                                                                       &                                                                                       \\ \cline{1-2}
\multicolumn{1}{l|}{\multirow{1}{*}{\texttt{gen} only}}                                                         & False                                                                                          & \begin{tabular}[c]{@{}r@{}}0.657\\ \small{($\pm$ 0.038)}\end{tabular}                           & \begin{tabular}[c]{@{}r@{}}0.528\\ \small{($\pm$ 0.027)}\end{tabular}                           & \begin{tabular}[c]{@{}r@{}}0.511\\ \small{($\pm$ 0.080)}\end{tabular}                          & \begin{tabular}[c]{@{}r@{}}0.420\\ \small{($\pm$ 0.104)}\end{tabular}                           & \begin{tabular}[c]{@{}r@{}}0.259\\ \small{($\pm$ 0.060)}\end{tabular}                            \\ \cline{1-2}

\multicolumn{1}{l|}{\multirow{1}{*}{(+) \texttt{clf}}}                                                          & False                                                                                          & \begin{tabular}[c]{@{}r@{}}0.571\\ \small{($\pm$ 0.110)}\end{tabular}                            & \begin{tabular}[c]{@{}r@{}}0.523\\ \small{($\pm$0.030)}\end{tabular}                             & \begin{tabular}[c]{@{}r@{}}0.489\\ \small{($\pm$ 0.115)}\end{tabular}                         & \begin{tabular}[c]{@{}r@{}}0.418\\ \small{($\pm$ 0.090)}\end{tabular}                            & \begin{tabular}[c]{@{}r@{}}0.283\\ \small{($\pm$ 0.070)}\end{tabular}                            \\ \cline{1-2}

\multicolumn{1}{l|}{\multirow{1}{*}{(+) \texttt{rank}}}                                                         & False                                                                                          & \begin{tabular}[c]{@{}r@{}}0.557\\ \small{($\pm$ 0.126)}\end{tabular}                           & \begin{tabular}[c]{@{}r@{}}0.518\\ \small{($\pm$ 0.031)}\end{tabular}                           & \begin{tabular}[c]{@{}r@{}}0.489\\ \small{($\pm$ 0.073)}\end{tabular}                         & \begin{tabular}[c]{@{}r@{}}0.436\\ \small{($\pm$ 0.089)}\end{tabular}                           & \begin{tabular}[c]{@{}r@{}}0.283\\ \small{($\pm$ 0.053)}\end{tabular}                           \\ \cline{1-2}

\multicolumn{1}{l|}{\multirow{2}{*}{\begin{tabular}[c]{@{}l@{}}(+) \texttt{rank}\\ $\quad$ + \texttt{clf}\end{tabular}}} & False                                                                                          & \begin{tabular}[c]{@{}r@{}}\underline{0.706}\\ \small{($\pm$ 0.001)}\end{tabular}                           & \begin{tabular}[c]{@{}r@{}}\textbf{0.559}\\ \small{($\pm$ 0.008)}\end{tabular} & \begin{tabular}[c]{@{}r@{}}\underline{0.656}\\ \small{($\pm$ 0.027)}\end{tabular}                         & \begin{tabular}[c]{@{}r@{}}\underline{0.546}\\ \small{($\pm$ 0.006)}\end{tabular}                           & \begin{tabular}[c]{@{}r@{}}\underline{0.371}\\ \small{($\pm$ 0.014)}\end{tabular}                           \\ \cline{2-2}

\multicolumn{1}{l|}{}                                                                                  & True                                                                                           & \begin{tabular}[c]{@{}r@{}}0.551\\ \small{($\pm$ 0.003)}\end{tabular}                           & \begin{tabular}[c]{@{}r@{}}0.473\\ \small{($\pm$ 0.000)}\end{tabular}                             & \begin{tabular}[c]{@{}r@{}}0.470\\ \small{($\pm$ 0.005)}\end{tabular}                         & \begin{tabular}[c]{@{}r@{}}0.324\\ \small{($\pm$ 0.010)}\end{tabular}                            & \begin{tabular}[c]{@{}r@{}}0.204\\ \small{($\pm$ 0.007)}\end{tabular}                           \\ \hline

\multicolumn{2}{c|}{\hl{\textbf{w/ ranking layer}}}                                                                                                                                             &                                                                                       &                                                                                       &                                                                                     &                                                                                       &                                                                                       \\
\multicolumn{1}{c}{\textbf{Input}}                                                                              & \textbf{Reasoning}                                                                                      &                                                                                       &                                                                                       &                                                                                     &                                                                                       &                                                                                       \\ \cline{1-2}
\multicolumn{1}{l|}{\multirow{3}{*}{\texttt{<|Resp.|>}}}    & False                                                                                          & \begin{tabular}[c]{@{}r@{}}0.624\\ \small{($\pm$ 0.136)}\end{tabular}                           & \begin{tabular}[c]{@{}r@{}}\underline{0.548}\\ \small{($\pm$ 0.033)}\end{tabular}                           & \begin{tabular}[c]{@{}r@{}}0.596\\ \small{($\pm$ 0.116)}\end{tabular}                         & \begin{tabular}[c]{@{}r@{}}0.485\\ \small{($\pm$ 0.049)}\end{tabular}                           & \begin{tabular}[c]{@{}r@{}}0.361\\ \small{($\pm$ 0.084)}\end{tabular}                           \\ \cline{2-2}
\multicolumn{1}{l|}{}                                                                                  & \begin{tabular}[c]{@{}c@{}}False\\ (\textcolor{red}{\texttt{rank only}})\end{tabular} & \begin{tabular}[c]{@{}r@{}}0.647\\ \small{($\pm$ 0.049)}\end{tabular}                           & \begin{tabular}[c]{@{}r@{}}0.533\\ \small{($\pm$ 0.024)}\end{tabular}                           & \begin{tabular}[c]{@{}r@{}}0.510\\ \small{($\pm$ 0.055)}\end{tabular}                          & \begin{tabular}[c]{@{}r@{}}0.431\\ \small{($\pm$ 0.058)}\end{tabular}                           & \begin{tabular}[c]{@{}r@{}}0.278\\ \small{($\pm$ 0.046)}\end{tabular}                           \\ \cline{2-2}
\multicolumn{1}{l|}{}                                                                                  & True                                                                                           & \begin{tabular}[c]{@{}r@{}}\textbf{0.735}\\ \small{($\pm$ 0.011)}\end{tabular} & \begin{tabular}[c]{@{}r@{}}\underline{0.548}\\ \small{($\pm$ 0.034)}\end{tabular}                           & \begin{tabular}[c]{@{}r@{}}\textbf{0.660}\\ \small{($\pm$ 0.030)}\end{tabular} & \begin{tabular}[c]{@{}r@{}}\textbf{0.548}\\ \small{($\pm$ 0.015)}\end{tabular} & \begin{tabular}[c]{@{}r@{}}\textbf{0.421}\\ \small{($\pm$ 0.007)}\end{tabular} \\ \cline{1-2}
\multicolumn{1}{l|}{\texttt{<|Rsn.|>}}                       & True                                                                                           & \begin{tabular}[c]{@{}r@{}} 0.650 \\ \small{($\pm$ 0.071)}\end{tabular} & \begin{tabular}[c]{@{}r@{}}0.540\\ \small{($\pm$ 0.032)}\end{tabular}                           & \begin{tabular}[c]{@{}r@{}}0.620\\ \small{($\pm$ 0.058)}\end{tabular} & \begin{tabular}[c]{@{}r@{}}0.446\\ \small{($\pm$ 0.071)}\end{tabular} & \begin{tabular}[c]{@{}r@{}}0.314\\ \small{($\pm$ 0.073)}\end{tabular} \\ \hline \hline
\end{tabular}

\end{adjustbox}
\end{table}

\footnotetext{The relevance score $s$, calculated without a ranking layer, is the same as \textbf{GPT} (vanilla), as detailed in \ref{appendix:baselines}.}

Moreover, we observed that jointly training the ranking layer alongside label and reasoning generation yields substantial advantages. Our best model significantly outperformed the \textcolor{red}{\texttt{rank only}} model shown in Table \ref{tab:gpt_performance}, which was solely trained to optimize relevance scores (\textcolor{red}{\texttt{Score}} in Figure \ref{fig:gpt}), without label and reasoning generation (\textcolor{red}{\texttt{Label}}, \textcolor{red}{\texttt{Reasoning}} in Figure \ref{fig:gpt}). This improvement shows the efficacy of our training approach, maintaining simplicity in inference. Furthermore, the ranking layer accelerates learning. Our best model converges in half the steps compared to the second-best model, which does not use a ranking layer. Specifically, the average training convergence steps were $5666.67(\pm 2624.67)$ and $11333.33 (\pm 1885.62)$, respectively.

\subsubsection{\textbf{Serving}} 
\label{sec:serving_exp}
\begin{figure}[h]
  \centering
  \includegraphics[width=\linewidth]{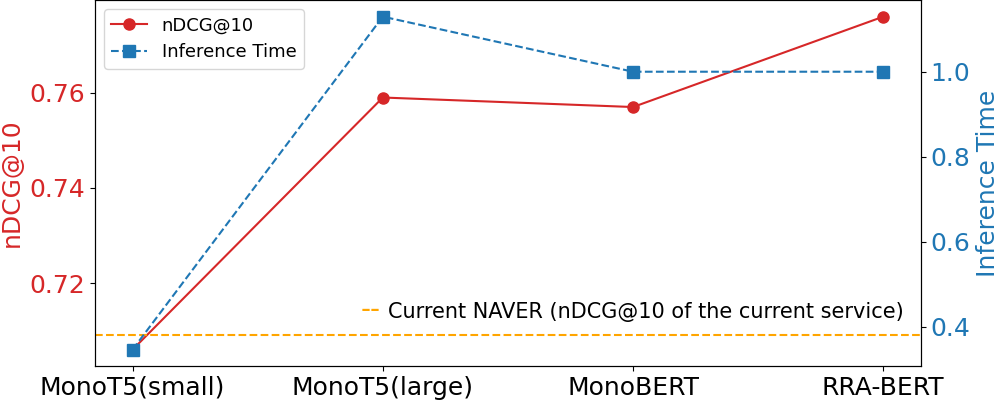}  
  \caption{Comparison of inference time (ratio) and nDCG@10 across four models. The yellow line represents the nDCG@10 performance of the current NAVER service.}
  \label{fig:serving}
\end{figure}

We compared response latency and ranking performance according to model types and sizes in Figure \ref{fig:serving}. All results were obtained using a single A100 GPU and model sizes follow the order: T5(small) < BERT < T5(large). As RRA-BERT shows the best ranking performance with reasonable speed we chose it as our final model and successfully deployed using TensorRT-LLM\footnote{https://github.com/NVIDIA/TensorRT-LLM} in real-world scenarios. More details are described in \ref{appendix:serving}.

\subsection{A/B Testing}
\label{sec:ab_test}
We conducted online and offline A/B tests comparing search results ranked by RRA-BERT with the current search results of NAVER search engine for long-tail queries. In the 7-day online A/B testing, RRA-BERT increased CTR by 5.63\%, top-1 document clicks by 5.9\%, and dwell time (the duration of time spending on a search result) by 7.97\%. Additionally, we conducted human evaluations by sampling 2,000 queries and scraping search results from both our ranker and the existing one. Human evaluators rated each search result on a scale of 1 to 5, while the ranker remains undisclosed. Compared to the existing ranker, ours achieved an average score increase of 1.41\%, with a significant 6.95\% boost for top-ranked documents. These results align with the findings of the online A/B tests, demonstrating our ranker's effectiveness in ranking relevant documents at the top of search results.

\section{Conclusion}
In this paper, we present an efficient label generation pipeline using LLMs that utilizes a pre-ranker to select effective documents and leverages LLMs' \textit{missing} phenomenon as a useful signal. We also present effective training methods to capture long and complex query-document relevance in both BERT and GPT. However, only key parts of the model structure are employed during inference, minimizing the service deployment burden. Through extensive experiments, including A/B testing on NAVER Search, we demonstrate the effectiveness of our approach for long-tail queries re-ranking. We believe that our approach, which captures context while focusing on key terms or features, will perform effectively across various domains, such as e-commerce \cite{lee2023mvfs, wang2022autofield, concf}. We expect that it will serve as a valuable direction for future applications in industry.  


\bibliography{custom}

\appendix
\section{Appendix}
\subsection{Details of Label Generation with LLMs}
\label{appendix:label_generation_detail}

\subsubsection{\textbf{Long-tail query sampling}}
\label{appendix:longtail_query}

\begin{figure}[h]
\centering  \includegraphics[width=\linewidth]{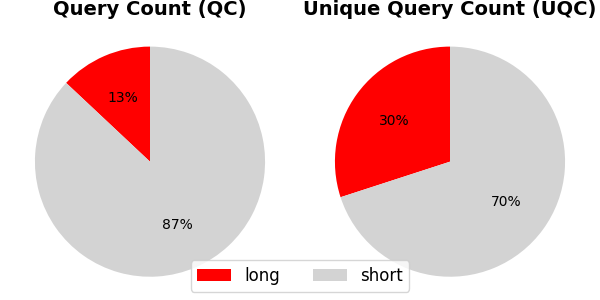}
  \caption{Proportion of long and short queries in our search engine's daily search logs. Queries with a length greater than $n$ are considered long, while others are considered short, based on internal criteria.}
  \label{fig:qc_uqc}
\end{figure}

\begin{figure}[h!]
\centering
\includegraphics[width=\linewidth]{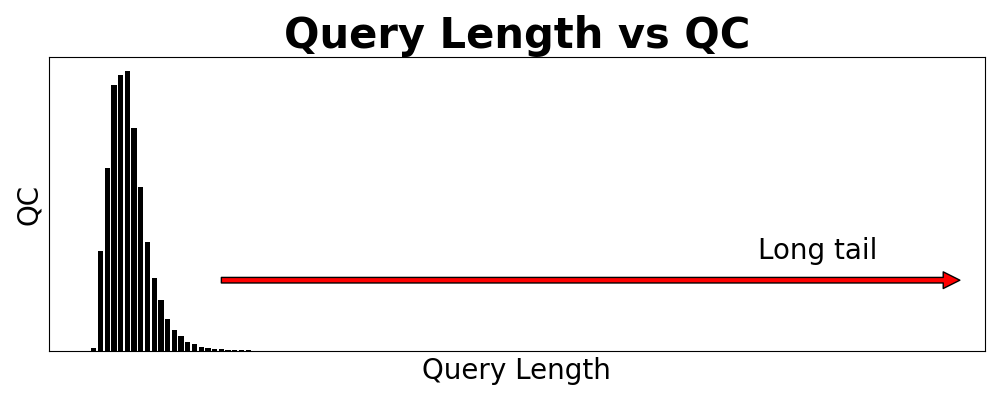}
  \caption{Aggregation of QC based on query lengths from daily search logs on our search engine, showcasing a long-tail distribution. Short queries with high QC are short-head, while long queries with low QC are long-tail.}
  \label{fig:qlen_qc}
\end{figure}

The NAVER search engine has predominantly handled short keyword-based queries, which constituted the majority of incoming queries as shown in Figure \ref{fig:qc_uqc}. Due to the high volume of popular short queries (short-head), ranking based on query-specific feedback features was effective, with feedback features often proving more significant than relevance features. However, this approach does not apply well to long-tail queries which are long and complex with lower user engagement which is depicted in Figure \ref{fig:qlen_qc}. Individual long-tail queries lacks sufficient user feedback, necessitating a reliance on relevance-based ranking. Since most of the incoming queries were short-head queries that are brief and simple, lacking contextual information, the combination of syntactic matching (like BM25) and feedback features is effective at scale. However, for long-tail queries, relevance is more effective than feedback and semantic matching is more effective than syntactic matching. This is because they lack user feedback but contain rich semantic information.

Now, with the capability of LLMs to understand the relevance between long and complex queries and documents, it has become possible to easily create high-quality training data targeted at long-tail queries. Through LLM distillation, it is now feasible to develop small semantic relevance models that mimic LLMs, thereby improving satisfaction with long-tail queries that were previously deemed less important. While individual long-tail queries may not be popular, relatively high UQC suggests that users have their own long-tail queries, making improvements to these query search results meaningful. To create specific training sets aligned with our objectives, we sampled queries that were lengthy, complex, infrequently searched based on internal criteria, corresponding to the tail part in Figure \ref{fig:qlen_qc}.

\subsubsection{\textbf{Pre-ranking}}
\label{appendix:preranking}
Pre-ranking is a crucial step in our ranking label generation pipeline, where all retrieved documents are ranked before being inputted into the LLM. Our pre-ranker, which is BERT-based, was trained on a small dataset of 1,000 queries and corresponding document snippets crawled from NAVER search pages, employing our label generation approach. As several studies  \cite{wang2023large, qin2023large, sun-etal-2023-chatgpt} have shown that the order of inputs significantly influences the performance of LLMs, we initially rank the documents using pre-ranker before inputting them into LLMs. Any rankers can be served as a pre-ranker, and the performance of ours can be found in Table \ref{tab:preranker}.

\begin{table}[h!]
\caption{The performance of our pre-ranker measured by nDCG@10}
\label{tab:preranker}
\centering
\begin{adjustbox}{width=\linewidth}
\begin{tabular}{c|c|c|c|c}
\hline
\textbf{NAVER}             & \textbf{\begin{tabular}[c]{@{}c@{}}MS\\ MARCO\end{tabular}} & \textbf{MIRACL}            & \textbf{DL19}              & \textbf{DL20}             \\ \hline
\multicolumn{1}{r|}{0.733} & \multicolumn{1}{r|}{0.567}                                  & \multicolumn{1}{r|}{0.653} & \multicolumn{1}{r|}{0.585} & \multicolumn{1}{r}{0.493} \\ \hline
\end{tabular}
\end{adjustbox}
\end{table}

\subsubsection{\textbf{Reasoning generation}}
\label{appendix:reasoning}
To train the GPT ranker, we utilized HCX-S to create reasoning, which explains the basis for considering $q$ and $d$ as either \textit{relevant} or \textit{irrelevant}.
\begin{equation}
\resizebox{\linewidth}{!}{$
\begin{aligned}
    label = \text{\textit{irrelevant}} \; \text{if} \; d \in D_{\text{excluded}} \; \text{else} \; \text{\textit{relevant}} \\
    reasoning_{(q,d,label)} = HCX_{\text{S}}(Prompt(q, d, label))
\end{aligned}
$}
\end{equation}
\noindent Generating reasoning along with HCX-L list-wise labeling results in either grouping multiple documents together for explanation or omitting reasoning for $D_{\text{excluded}}$. Therefore, we opted for point-wise reasoning generation using HCX-S, a small model of HCX that provides adequate reasoning. Here we utilized a simple prompt like "Explain why the given $q$ and $d$ pair is classified as $label$ (= \textit{relevant} or \textit{irrelevant})."

\subsection{Experimental settings}
\label{appendix:settings}
\subsubsection{\textbf{Datasets \& Metrics}}
\label{appendix:dataset_and_metrics}
The method of constructing training data is in Section \ref{sec:label_generation}. For evaluation, we set aside 10\% of this data. To ensure the robust performance of our models, we also evaluated on public testsets for passage re-ranking: MS MARCO \cite{bajaj2018ms}, MIRACL \cite{10.1162/tacl_a_00595}, DL19 \cite{DBLP:journals/corr/abs-2003-07820}, and DL20 \cite{DBLP:journals/corr/abs-2102-07662}. The statistics of testsets are in Table \ref{tab:dataset}. Since MS MARCO, DL19, and DL20 are in English, we translated them into Korean using NAVER Papago\footnote{https://papago.naver.com/}, which is a Korean machine translation service provided by NAVER. MIRACL is a multilingual dataset, so we directly used its Korean version. We used nDCG \cite{10.1145/582415.582418} as the evaluation metric, widely used for measuring ranking performance.

\begin{table}[h!]
\caption{The number of queries and documents in five testsets.}
\label{tab:dataset}
\centering
\begin{adjustbox}{width=\linewidth}

\begin{tabular}{l|r|r|r|r|r}
\hline
\textbf{Count}           & \multicolumn{1}{c|}{\textbf{Ours}} & \multicolumn{1}{c|}{\textbf{\begin{tabular}[c]{@{}c@{}}MS\\ MARCO\end{tabular}}} & \multicolumn{1}{c|}{\textbf{MIRACL}} & \multicolumn{1}{c|}{\textbf{DL19}} & \multicolumn{1}{c}{\textbf{DL20}} \\ \hline

\textbf{Query}           & 700                               & 9,650                                                                            & 213                                  & 43                                 & 54                                \\
\textbf{Document} (avg.) & 20                                  & 8.20                                                                             & 14.35                                & 107.26                             & 105.30                            \\ \hline
\end{tabular}

\end{adjustbox}
\end{table}

\subsubsection{\textbf{Baselines}}
\label{appendix:baselines}
We include the following baselines. MonoBERT and MonoT5 are trained on the dataset created by our label generation pipeline. For RankGPT training labels, we employed sliding windows on $D$, following the paper \cite{sun-etal-2023-chatgpt}, with a window size of 20 and a step size of 10, without our pre-ranking process.

\begin{itemize}
    \item \textbf{HCX-L} \cite{yoo2024hyperclova}: With our teacher model HCX-L which is an instruction-tuned LLM including diverse Korean data, we utilized the list-wise ranking prompt from RankGPT \cite{sun-etal-2023-chatgpt}.
    \item \textbf{BERT} (naive): Before tuning our backbone BERT, we rank based on the cosine similarity of embeddings for $q$ and $d$.
    \item \textbf{GPT} (vanilla): Before tuning our backbone GPT, we rank using the relevance score  $p_{\text{rel}} - p_{\text{irrel}}$ (see Equation \ref{eq:gpt_rel_score}, changing \texttt{<|Relevant|>} to \textit{Relevant} and to \texttt{<|Irrelevant|>} to \textit{Irrelevant}).
    \item \textbf{BM25} \cite{robertson2009probabilisticbm25}: A probabilistic score for $(q,d)$, calculated from term frequency and inverse document frequency.
    \item \textbf{MonoBERT} \cite{nogueira2019passage}: A point-wise trained classification model for determining $(q,d)$ relevance, based on BERT.
    \item \textbf{MonoT5} \cite{nogueira2020document}: A T5-based ranker, also employing a classification based approach.
    \item \textbf{RankGPT} \cite{sun-etal-2023-chatgpt}: BERT and GPT-based ranker trained in a pair-wise manner to mimic list-wise LLM ranking.
\end{itemize}

\begin{table*}[t!]
\renewcommand{\arraystretch}{1.5}
\centering
\small
\begin{tabular}{|p{0.05\linewidth}|p{0.43\linewidth}|p{0.43\linewidth}|}
\hline
\multicolumn{1}{|c|}{} & \multicolumn{2}{c|}{\textbf{Query}} \\ \hline
 & \textbf{8개월 강아지가 잠\textcolor{red}{만}자요} \textcolor{gray}{\textit{(8-month-old dog \textcolor{red}{only} sleeps)}} & \textbf{8개월 강아지가 잠\textcolor{red}{안}자요} \textcolor{gray}{\textit{(8-month-old dog \textcolor{red}{doesn't} sleep)}} \\ \hline
\multicolumn{1}{|c|}{\textbf{Rank}} & \multicolumn{2}{c|}{\textbf{Document}} \\ \hline
\centering \textbf{1} & [갑자기 잠만 자는 강아지?] 8개월 비숑인데요... 성격 활발하고 집에서도 뛰어다니는데 며칠 전부터 계속 잠만 자네요... 
\newline \textcolor{gray}{\textit{[Suddenly sleeping all the time?] I have an 8-month-old Bichon... It's very active but has been sleeping constantly for a few days...}} & 
[우리 강아지가 밤에 안자요] 강아지 8개월인데, 저녁에 계속 놀아달라고 하고 너무 활발해요...
\newline \textcolor{gray}{\textit{[Our dog doesn't sleep at night] Our 8-month-old dog keeps playing all evening and is super energetic...}} \\ \hline
\centering \textbf{2} & [강아지가 밥도 안 먹고 잠만 자요] 강아지가 밥도 안 먹고 잠을 많이 자는데, 어디 아픈 걸까요? 치와와이고 8개월 되었어요... 
\newline \textcolor{gray}{\textit{[Dog isn't eating and only sleeping] My 8-month-old Chihuahua isn't eating and sleeps all the time. Could it be sick?}} & 
[강아지가 밤에 잠을 안자요! 원인과 대처법] 강아지가 밤새도록 울면 보호자도 지치기 쉬운데...
\newline \textcolor{gray}{\textit{[The dog isn't sleeping at night! Causes and solutions] If a dog cries all night, it can exhaust its owner...}} \\ \hline
\centering \textbf{3} & [8개월 강아지 잠이 원래 이렇게 많나요?] 하루 20시간씩 자요... 스트레스 때문일까요? 병원 다녀왔는데...
\newline \textcolor{gray}{\textit{[Is it normal for an 8-month-old dog to sleep this much?] It sleeps 20 hours a day... Could it be stress? We've visited the vet...}} & 
[새끼 강아지가 잠을 안자요] 새끼 강아지가 밤에 자지 않아서 보호자의 수면을 방해해요...
\newline \textcolor{gray}{\textit{[Puppy not sleeping] My puppy doesn't sleep at night and disturbs the owner's sleep...}} \\ \hline
\centering \textbf{4} & [강아지가 잠만 자는 이유가 뭘까요?] 강아지가 하루 종일 잠만 자는데, 어디 아픈 걸까요? 밥도 잘 안 먹고...
\newline \textcolor{gray}{\textit{[Why is my dog only sleeping?] My dog is sleeping all day long. Could it be sick? It doesn't eat well either...}} & 
[아기 강아지 수면시간] 강아지가 잠을 너무 많이 자거나 너무 적게 자서 보호자가 걱정할 수 있어요...
\newline \textcolor{gray}{\textit{[Puppy sleep time] A puppy sleeping too much or too little can make the owner worry...}} \\ \hline
\end{tabular}
\caption{Document ranking results for two queries with a one-character difference that completely changes their meaning, using our RRA-BERT model.}
\label{tab:example_1}
\end{table*}

\subsubsection{\textbf{Implementation details}}
\label{appendix:implementation_details}
We used internally pre-trained small Korean LMs, specifically RoBERTa \cite{liu2019roberta} for RRA-BERT and GPT \cite{gpt2} for RRA-GPT. The same backbones were also used for MonoBERT and RankGPT in baseline experiments. Additionally, we used an in-house T5 backbone for MonoT5. For hyperparameters, the optimizer used for all models is AdamW \cite{loshchilov2019decoupled}, with a learning rate of 1e-5. We trained each model three times and reported the averaged performance. Training data was randomly split into a 9:1 ratio for training and validation. BERT was validated every 300 steps, GPT every 1,000 steps. Early stopping occurred if performance didn't improve for five consecutive validations. The generated ranking labels were used in training as follows: 
\begin{equation}
\resizebox{\linewidth}{!}{$
\begin{aligned}
2-i*0.1 \; \text{for} \; d_i \in D_{\text{ranked}} \\
0.2 - (j+1)*0.01 \; \text{for} \; d_j \in D_{\text{excluded}} \\
0 \; \text{for} \; d_{\text{neg}}; \; \text{where $d_{\text{neg}}$ are negatives with a size of 3}
\end{aligned}
$}
\end{equation}
\noindent To clarify, $j$ is randomly assigned, since there is no order in $D_{\text{excluded}}$. For MonoBERT and MonoT5 training, which utilize a classification task framework, we label $D_{ranked}$ as relevant documents and $D_{excluded}$ as irrelevant ones. In addition, for TCL training of RRA-BERT, we set $k$, the number of selected document tokens per query token, to 3 for token selection (Section \ref{sec:token_selection}), and $\alpha$ to 0.3 as the hyperparameter for combining TCL loss (Section \ref{sec:optimization}).

\subsection{\textbf{Serving details}}
\label{appendix:serving}
First of all, despite using a small-sized GPT, its speed and throughput were not as good as BERT and T5, making it difficult to use it for search engine, which demand high Queries Per Second (QPS). With RRA-BERT, we measured performance degradation and QPS based on floating point format. Compared to float32 used during training, using bfloat16, performance was 99.8\%, and with fp8, it was 99\%, both offering an 18\% QPS advantage. Consequently, we selected bfloat16 for serving and successfully deployed to the service.

\subsection{\textbf{Real-world qualitative examples}}
\label{appendix:example}
We provide a few examples of long-tail queries ranked by our final model, RRA-BERT. The original query and document were in Korean and have been translated into English; the document refers to the [title] and snippet. Table \ref{tab:example_1} illustrates two queries where a single character change completely alters the meaning: "8개월 강아지가 잠\textcolor{red}{만}자요" \textcolor{gray}{\textit{(8-month-old dog \textcolor{red}{only} sleeps)}}" vs "8개월 강아지가 잠\textcolor{red}{안}자요" \textcolor{gray}{\textit{(8-month-old dog \textcolor{red}{doesn't} sleep)}}. This example shows how RRA-BERT successfully distinguishes between the two semantically different queries despite their minimal character variation. Additionally, Table \ref{tab:example_2}, \ref{tab:example_3} and \ref{tab:example_4} present ranking results for single long-tail queries, demonstrating how our model ranks relevant documents at the top by capturing important term signals while still considering the overall semantic context.

\begin{table*}[t!]
\renewcommand{\arraystretch}{1.5} 
\centering
\small
\begin{tabular}{|p{0.1\linewidth}|p{0.85\linewidth}|} 
\hline
\multicolumn{2}{|c|}{\shortstack{\\[0.2ex] \textbf{Query = 80대 요관암 말기 암 항암치료} \\[0.1ex] \textcolor{gray}{\textit{(Chemotherapy treatment for terminal stage ureteral cancer in people in their 80s)}}}} \\ \hline
\multicolumn{1}{|c|}{\textbf{Rank}} & \multicolumn{1}{c|}{\textbf{Document}} \\ \hline
\centering \textbf{1} & [아버지가 요관암 말기 판정 받으셨어요] 아버지가 올해 만 71세이신데 요관암 말기 판정 받으셨어요. 참전용사이셔서 집근처 중앙보훈병원에서 검사받고 진료 받으셨는데 뼈와 임파선까지 전이가 되어 수술이 불가하다고 하여 항암치료 상담하려고 3월 26일 혈액종양과 진료 예약해놓은 상태인데 차병원 혈종과가 잘 봐주신다고 하여 차병원 천재경 교수님 진료를 3월19일 예약해놓은 상황이에요. 근데 제가… 
\newline \textcolor{gray}{\textit{[My father was diagnosed with terminal ureter cancer] My father is 71 years old this year and was diagnosed with terminal ureter cancer. He is a war veteran and received tests and care at the Central Veterans Hospital near our house. However, since the cancer has spread to the bones and lymph nodes, surgery is not an option. We have scheduled a consultation for chemotherapy on March 26th with the hematology-oncology department, and we also made an appointment with Dr. Cheon Jae-kyung at Cha Hospital for March 19th for another consultation... But I...}} \\ \hline
\centering \textbf{2} & [아버지 요관암 말기] 저희 친정 아버지가 2주전 요관암 말기 진단을 받으셨어요. 지방 대학병에서는 항암도 어렵다하여 신촌세브란스 최영득교수한테 며칠전 진료를 받았어요. 진료들어가기전에 서브... 정말 너무 화가 치밀었지만 항암치료라도 해보자고 하니 을이된 입장으로 더 이상 말도 못하고 일단 돌아왔습니다. 이런 의사한테 아버지 치료를 맡겨야 하는건지 모르겠습니다. 
\newline \textcolor{gray}{\textit{[Father’s terminal ureter cancer] My father was diagnosed with terminal ureter cancer two weeks ago. At a local university hospital, they said chemotherapy would be difficult, so we saw Dr. Choi Young-deuk at Sinchon Severance a few days ago. Before going into the consultation... I was really furious, but when they suggested trying chemotherapy, I felt like a subordinate and couldn't say anything more, so we just left. I don't know if I should trust this doctor with my father’s treatment.}} \\ \hline
\centering \textbf{3} & [요관암 4기 말기 원인과 증상, 생존율 알아보기 (부작용/항암치료방법/병원)] 원인, 생존율, 요관암 4기 치료 방법, 요관암 말기병원에 대해 알아보는 시간을 가졌습니다. 전체 암 발생 중 1\% 미만에 해당되는 만큼 요관암 말기에 발견되는 경우가 많습니다. 하지만 포기하지 않고 신체 상태와 요관암 항암치료를 꾸준히 이행하면서 요관암 말기 병원에서 후유증 완화를 위한 면역 관리에 꾸준히 노력하신다면 충분히 암을 이겨내실 수 있을 것입니다. 
\newline \textcolor{gray}{\textit{[Understanding Stage 4 Ureter Cancer: Causes, Symptoms, Survival Rates (Side Effects, Chemotherapy Methods, Hospitals)] We took some time to understand the causes, survival rates, stage 4 ureter cancer treatments, and hospitals for terminal ureter cancer. Since ureter cancer accounts for less than 1\% of all cancer cases, it is often detected in its terminal stage. However, if you keep up with chemotherapy and consistent immune management for side-effect relief, there's a good chance you can overcome the cancer.}} \\ \hline
\centering \textbf{4} & [아빠의 투병 일지] 2022년 1월 아빠는 요관암 수술을 받았다. 80대인 아빠는 항암을 거부하셨고 3개월에 한번 검사를 받았다. 2023년 6월 정기검사에서 재발과 전이가 되었다. 아빠에게 재발과 전이 됐다는 소식을 전했다. 이제 철드니 부모님이 너무 아프시고 늙어 계신다. 오늘 아빠 항암 맞으러 가는 엘리베이터에서 함께 찍은 사진을 남겼다. 80대 부모님, 엄마, 아빠, 암투병, 항암치료, 힘내자. 
\newline \textcolor{gray}{\textit{[Dad's Cancer Battle Journal] In January 2022, my father had surgery for ureter cancer. At 80 years old, my father refused chemotherapy and underwent tests every three months. In June 2023, during a routine exam, we found out the cancer had recurred and spread. I broke the news to my father that it had returned and metastasized. Now that I'm finally growing up, my parents are so sick and frail. Today, we took a photo in the elevator while taking my dad to chemotherapy. My 80-year-old parents, mom, dad, cancer battle, chemotherapy, stay strong.}} \\ \hline
\end{tabular}
\caption{Document ranking results for the long-tail query \textbf{"80대 요관암 말기 암 항암치료"} \textcolor{gray}{\textit{(Chemotherapy treatment for terminal stage ureteral cancer in people in their 80s)}}. This query includes a highly specific intent, targeting a particular type of cancer, specific stage and a specific age group. The results must focus on this precise scenario, and not be confused with treatments for other cancers or stage or age groups. Our method ensures that documents related to the treatment process, recurrence, questions for elderly patients with ureteral cancer are ranked higher.}
\label{tab:example_2}
\end{table*}

\begin{table*}[ht]
\renewcommand{\arraystretch}{1.5} 
\centering
\small
\begin{tabular}{|p{0.1\linewidth}|p{0.85\linewidth}|} 
\hline
\multicolumn{2}{|c|}{\shortstack{\\[0.2ex] \textbf{Query = 흑백요리사 11화 넷플릭스에 올라오는 요일과 시간} \\[0.1ex] \textcolor{gray}{\textit{(Episode 11 release day and time for Culinary Class Wars on Netflix)}}}} \\ \hline
\multicolumn{1}{|c|}{\textbf{Rank}} & \multicolumn{1}{c|}{\textbf{Document}} \\ \hline
\centering \textbf{1} & [넷플릭스 흑백요리사 11화 예고∥업로드 공개 시간] ▶흑백 식당 예약◀ 흑백요리사 예고 공개시간(업로드) 선공개 매주 화요일 오후 4시 1화, 2화,3화, 4화 2024년 9월 17일(화) 5화, 6화, 7화 2024년 9월 24일(화) 8화, 9화, 10화 2024년 10월 1일(화) 11화, 12화 2024년 10월 8일(화) ‘흑백요리사’는 넷플릭스 글로벌 TOP 10 TV(비영어) 부문에서 1위를 차지했습니다... 
\newline \textcolor{gray}{\textit{[Netflix Culinary Class Wars Episode 11 Preview∥Upload Release Time] ▶Culinary Class Wars Restaurant Reservation◀ Culinary Class Wars preview release time (upload) early release every Tuesday at 4 p.m. Episodes 1, 2, 3, 4 on September 17, 2024 (Tue), Episodes 5, 6, 7 on September 24, 2024 (Tue), Episodes 8, 9, 10 on October 1, 2024 (Tue), Episodes 11, 12 on October 8, 2024 (Tue). 'Culinary Class Wars' ranked 1st on Netflix's global TOP 10 TV (non-English) section...}} \\ \hline
\centering \textbf{2} & [넷플릭스 흑백요리사 11화 예고∥업로드 공개 시간 - KakaoNaver] 누구도 그들의 날카로운 심사를 피해갈 순 없다. <흑백요리사: 요리 계급 전쟁>, 지금 오직 넷플릭스에서. 일반적으로 넷플릭스 공식 SNS 채널이나 유튜브 채널에서 예고편을 공개하는 경우가 많으므로, 해당 플랫폼들을 확인해보시는 것이 좋을 것 같습니다. 
\newline \textcolor{gray}{\textit{[Netflix Culinary Class Wars Episode 11 Preview∥Upload Release Time - KakaoNaver] No one can escape their sharp judgment. <Black and White Chef: Culinary Class Wars>, now only on Netflix. Previews are often released on Netflix's official social media or YouTube channels, so it’s a good idea to check those platforms.}} \\ \hline
\centering \textbf{3} & [넷플릭스 흑백요리사 공개시간 심사위원은 누구?] 공개시간? 나는 화요일날 땡치면 올라오는지 알았지? 흑백요리사 첫화 방영 이후 다음화는 언제 업데이트 되나 그게 제일 궁금 하더라구요 넷플릭스에서 보니 화요일마다 조금씩... 알아보니 흑백요리사의 정확한 업데이트 시간은 '화요일 오후4시' 라고 하네요 지금은 1~10화까지 공개되어있으며 마지막인 11화, 12화는 10월 8일 오후 4시에 공개가 될 것 같습니다 우승자는... 
\newline \textcolor{gray}{\textit{[Netflix Culinary Class Wars Release Time, Who Are the Judges?] Release time? I thought it would be uploaded exactly on Tuesday. After the airing of the first episode of Culinary Class Wars, I was most curious about when the next episodes would be updated. I found out that Culinary Class Wars is updated 'every Tuesday at 4 p.m.' Right now, episodes 1 to 10 are available, and the last episodes, 11 and 12, will be released on October 8 at 4 p.m. The winner is...}} \\ \hline
\centering \textbf{4} & [넷플릭스 흑백요리사 공개시간 몇부작 제작사 백종원 등 출연진 정보] 총 20명의 유명 요리사 '백수저' 셰프들과 80명의 흑수저 셰프가 출연해 요리 경연을 펼칩니다. 흑백요리사 공개시간 2024년 9월 17일 화요일 오후 4시에 넷플릭스(Netflix)를 통해서만 공개됩니다. 한 번에 전편이 모두 공개되는 넷플릭스 드라마와는 다르게, 9월 17일 화요일 오후 4시에 1회부터 4회까지 한번에 공개되고, 매 주 순차적으로 나머지 회차가 공개됩니다. 몇부작… 
\newline \textcolor{gray}{\textit{[Netflix Culinary Class Wars Release Time, Number of Episodes, Producers, and Cast Info] A total of 20 famous 'White Spoon' chefs and 80 'Black Spoon' chefs compete in cooking contests. Culinary Class Wars release time: 4 p.m. on Tuesday, September 17, 2024, exclusively on Netflix. Unlike some Netflix series that drop all episodes at once, Culinary Class Wars releases episodes in parts, with episodes 1 to 4 released on September 17, and the rest following weekly.}} \\ \hline
\end{tabular}
\caption{Document ranking results for the long-tail query \textbf{"흑백요리사 11화 넷플릭스에 올라오는 요일과 시간"} \textcolor{gray}{\textit{(Episode 11 release day and time for Culinary Class Wars on Netflix)}}. This query specifically targets the release schedule for episode 11 of a TV show. Our method accurately captures the relevant documents, focusing on precise information about release dates and times, and ranking those that explicitly mention episode details and related content higher, while not missing the overall semantic context.}
\label{tab:example_3}
\end{table*}

\begin{table*}[ht]
\renewcommand{\arraystretch}{1.5} 
\centering
\small
\begin{tabular}{|p{0.1\linewidth}|p{0.85\linewidth}|} 
\hline
\multicolumn{2}{|c|}{\shortstack{\\[0.2ex] \textbf{Query = 40대 실비 암보험 리모델링하는 방법} \\[0.1ex] \textcolor{gray}{\textit{(How to Remodel Cancer and Medical Insurance for People in Their 40s)}}}} \\ \hline
\multicolumn{1}{|c|}{\textbf{Rank}} & \multicolumn{1}{c|}{\textbf{Document}} \\ \hline
\centering \textbf{1} & [40대 보험 실비 암보험 리모델링] .. 지금 가지고 있는 병력은 없고, 40대 초반 기준으로 리모델링 하여 실비 + 암보험 한달 보험료는 얼마정도가 적절할까요? 혼자서 보험 가입에... 현실적으로 저축액을 정하시고 남는 금액으로 보험을 유지하는게 가장 좋은 방법이에요. 이번에 보험료 줄이기 꼭 성공하세요. 일단 필수적인... 
\newline \textcolor{gray}{\textit{[Cancer and Medical Expense Insurance Remodeling for People in Their 40s] ... There are no pre-existing conditions, and based on early 40s, how much would be appropriate for a monthly premium for remodeled cancer + medical expense insurance? ... Realistically, the best way is to set aside a savings amount and maintain the insurance with the remaining amount. Make sure to succeed in reducing your premium this time...}} \\ \hline
\centering \textbf{2} & [40대 의료실비보험 리모델링 해지] 40대 초반 공무원 입니다. 의료실비보험 암보험 리모델링 가입때문에 알아보고 있습니다. 지인에게 가입해 9년 정도 유지했던 한화생명... 이런 종합보험은 지금 해지하는것이 피해를 최소화 할 수 있는 방법이며, 다른 상품이나 같은회사의 상품으로 제대로 재설계, 리모델링을 권장하며...
\newline \textcolor{gray}{\textit{[Canceling and Remodeling Medical and Cancer Insurance in Their 40s] I’m in my early 40s and a civil servant. I’m looking into remodeling my medical and cancer insurance. I had been maintaining a policy with Hanwha Life for about 9 years... Canceling this type of comprehensive insurance now would minimize damages, and I recommend switching to a properly redesigned plan, even with the same company...}} \\ \hline
\centering \textbf{3} & [40대 여성 실비+암보험 어떻게 준비해야할까? :: 더 좋은생각] 실비+암 상담 및 가격/격적 무료조회 하기 ◆ 40대 여자 실비 + 암 보장은 월납입료가 얼마가 적당할까? 이는 정답이 없습니다. 그렇기 때문에 장기간 유지할 수 있는 정도로 부담없는 가격대로 준비하는 것이 유리할 수 있습니다. 장기적으로 가입을 유지하고 평생을 보장받는다면 비갱신형+순수보장형으로 설계를 하는 것이 비용을 줄일 수 있는 방법일 수 있습니다.... 
\newline \textcolor{gray}{\textit{[How Should Women in Their 40s Prepare for Medical + Cancer Insurance? :: Better Thoughts] Medical + cancer insurance consultation and free quotes available ◆ How much is appropriate for the monthly premium for a woman in her 40s? There is no correct answer. It’s best to set a premium at a reasonable price you can maintain for a long period. For long-term coverage, choosing a non-renewable, pure insurance plan is a cost-effective way to prepare...}} \\ \hline
\centering \textbf{4} & [40대 암보험] 40대 초반이고 실비만 1개있고 암보험이 없어서 가입해야 할 것 같아서 고민중입니다. 보장금액이나 항목을 어떻게 하면 좋을까요?
\newline \textcolor{gray}{\textit{[Cancer Insurance for People in Their 40s] I'm in my early 40s, and I only have one medical expense insurance plan, so I’m thinking of getting cancer insurance. What kind of coverage and terms would be best?}} \\ \hline
\end{tabular}
\caption{Document ranking results for the long-tail query \textbf{"40대 실비 암보험 리모델링하는 방법"} \textcolor{gray}{\textit{(How to Remodel Cancer and Medical Insurance for People in Their 40s)}}. Our method accurately captures both the specific age group and the detailed types of insurance, ranking documents with relevant advice on remodeling cancer and medical insurance higher.}
\label{tab:example_4}
\end{table*}

\end{document}